\documentclass[aps,prl,showpacs,twocolumn]{revtex4}
\usepackage{graphicx}
\begin{document}
\newcommand {\be}{\begin{equation}}
\newcommand {\ee}{\end{equation}}
\newcommand {\bea}{\begin{eqnarray}}
\newcommand {\eea}{\end{eqnarray}}
\newcommand {\nn}{\nonumber}
\newcommand {\siz}{\sum_{i=1}^{\zeta}}


\title{Inhomogeneous quantum antiferromagnetism on periodic lattices}

\author{  A. Jagannathan}
\affiliation{Laboratoire de Physique des Solides, CNRS-UMR 8502, Universit\'e Paris-Sud, 91405 Orsay, France }
\author{R. Moessner}
\affiliation{Laboratoire de Physique Th\'eorique de l'Ecole Normale Sup\'erieure, CNRS-UMR 8549, Paris, France}
\author{ S. Wessel}
\affiliation{Institut f\"ur Theoretische Physik III, Universit\"at Stuttgart, 70550 Stuttgart, Germany}

\date{\today}

\begin{abstract}
We study quantum antiferromagnets on two-dimensional bipartite
lattices. We focus on {\em local} variations in the properties of
the ordered phase which arise due to the presence of inequivalent
sites or bonds in the lattice structure, using linear spin wave
theory and quantum Monte Carlo methods.
Our primary finding is that sites with a
{\em high} coordination tend to have a {\em low} ordered moment,
at odds with the simple intuition of high coordination favoring
more robust N\'eel ordering. The lattices considered
are the dice lattice, which is dual to the kagome, the CaVO lattice,
an Archimedean lattice with two inequivalent bonds, and  the
crown lattice, a tiling of squares and rhombi
with a greater variety
of local environments.
We present results for the onsite
magnetizations and local bond expectation values for the spin-1/2
Heisenberg model on these lattices, and discuss the exactly
soluble model of a Heisenberg star, which provides a simple
analytical framework for understanding our lattice studies.

\end{abstract}
\maketitle

\section{\label{sec:level1}I. Introduction}
The long-wavelength universal properties of antiferromagnets on
bipartite lattices are well established \cite{chaikinlubensky}. In
this paper, we ask to what extent local properties -- the
dependence of the ordered moment or bond strength on local
coordination -- display systematic behavior. To do this, we consider
quantum Heisenberg antiferromagnets on various two-dimensional
lattices. These lattices are bipartite, so that classical 
antiferromagnets form collinear N\'eel states at zero
temperature; for quantum spins, this ordering persists, albeit with
reduction of the order parameter due to quantum fluctuations. A review
of results obtained for uniform Archimedean lattices (in which all
sites -- but not necessarily all bonds -- are equivalent) is given in
Ref.~\cite{richt}. Generally, the size of the order parameter
depends on dimensionality -- Heisenberg N\'eel order is absent in $d=1$
-- and coordination, $z$: it is bigger for the square ($z=4$) than for
the honeycomb ($z=3$) lattice.


Upon increasing the coordination $z$ the exchange `mean-field' acting on a
spin at a given lattice site grows. The larger this mean-field, the
less effective quantum fluctuations are in reducing the staggered
alignment of the spin at that site.  For inhomogeneous lattices with
sites of different coordination, this suggests that
the local staggered magnetization should {\it increase} with sites 
coordination $z$. Analogously, one would expect average bond
energies to decrease with $z$, as fluctuations transverse to the
ordered moment become increasingly difficult to coordinate between a
growing number of neighbors.

However, recent investigations on a two-dimensional, quasiperiodic
system, the octagonal tiling, with six different local coordinations,
$z=3,4,5,6,7,8$ lead to the result that the local staggered
magnetization does not follow this expectation; instead, 
both the local staggered
magnetization $m_i$ and the averaged local bond
strength of all bonds connected to a given site, $\overline{b}_i$,
tend to {\it decrease} with increasing $z$ (cf. Sec.~II for a formal
definition of both $m_i$ and $\overline{b}_i$). This property has been observed in quantum 
Monte Carlo simulations~\cite{wess1}, in linear spin wave theory~\cite{wess2} and
in a real space renormalization group approach based on the self-similar
structure of the quasiperiod lattice~\cite{jag}.
The absence of translation invariance, however, has important
consequences for the spectrum and eigenmodes of the spin
Hamiltonian~\cite{wess2}, and it is the invariance under scale
transformation that determines the real space properties of the
antiferromagnet~\cite{wess1,jag}. In order to assess the relevance of
the absence of translation symmetry on the behavior of the local order
parameter distribution, here we consider various {\em inhomogeneous but
periodic} lattices. 

In particular, we calculate the variations in
$m_i$ and $\overline{b}_i$ for lattices with small unit cells that
contain inequivalent bonds and/or sites. We consider the dice lattice,
the CaVO lattice and a lattice that to our knowledge has not been
studied previously, the crown lattice, which is built from squares and
rhombi. This structure shows similar local environments to those found
in the octagonal tiling, but unlike the latter possesses translational
invariance.

Our results on regular lattices are in qualitative agreement with
those of the octagonal tiling. To provide a simple analytical
explanation of the $z$-dependence of the local quantities, we appeal
to the exactly soluble case of the Heisenberg star
\cite{richterstar}. In particular, this resolves the conundrum of
a decreasing $m_i(z)$ by showing that, while the {\em average} moment
$m_i^{av}$ increases with the average coordination $z^{av}$ of the
lattice, an order-by-disorder type mechanism  concentrates the
reduction of the ordered moment to highly coordinated sites.

\section{\label{sec:levelmethod} II. Model and Numerical Methods}
In the following we consider the nearest neighbor Heisenberg
antiferromagnet, described by the  Hamiltonian
\begin{eqnarray}
 H = J \sum_{\langle i,j\rangle} \vec{S}_i \vec{S}_j
\end{eqnarray}
where the sum is taken over all bonds in the lattice.
$J>0$ is assumed to be independent of the local
environments, and set equal to 1 in the rest of the paper.
We employ a combination of linear spin wave theory~\cite{white} and quantum Monte Carlo simulations in
our analysis. Details of the linear spin wave approach will be presented in the following section for each specific lattice.

The quantum Monte Carlo simulations were performed using the
stochastic series expansion method~\cite{sse} for finite lattices.  In
particular, for the dice lattice we considered systems with $N_s=$
192, 432 and 768 sites, for the CaVO lattice those with 64, 144, 265
and 748 sites, and for the crown lattice, systems with 112, 252, 448
and 700 sites. In each case the temperature was taken sufficiently low to
obtain ground state expectation values. For a given site $i$ of the
finite size lattice, the local value of the staggered magnetization is
given by \bea m_i=\sqrt{\frac{3}{N_i}\sum_{j=1}^{N_i} (-1)^{i+j}
\langle S^z_i S^z_j \rangle}, \eea where the sum extends over all the $N_i$ 
lattice sites $j$ which are equivalent to site $i$ with respect to the antiferromagnetic unit-cell.
Finally, standard finite size scaling analysis was performed to obtain the
local staggered magnetization in the thermodynamic
limit~\cite{wess1}. For a given bond $\langle i,j
\rangle$ on the lattice, we define the corresponding bond strength
as $b_{i,j}=| \langle \vec{S}_i \vec{S}_j \rangle |$. Similarly, the
averaged local bond strength at a site $i$ is \bea \overline{b}_i
=  \frac{1}{z}\sum_{j=1}^{z} | \langle \vec{S}_i \vec{S}_j \rangle
|, \eea where $j$ extends over all the $z$ sizes that are connected to
site $i$~\cite{wess1}. In the following we present the results of
the quantum Monte Carlo simulations as well as those from linear
spin wave theory for the different lattices.

\section{\label{sec:level1a} III. Local properties of two-dimensional
bipartite lattices}

\subsection{\label{sec:level2} III.A The Dice lattice}
\begin{figure}[ht]
\begin{center}
\includegraphics[scale=0.5]{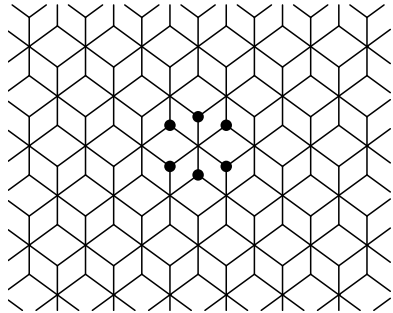}
\vspace{.2cm} \caption{The dice lattice, showing a localized excitation.}

\label{dice.fig}
\end{center}
\end{figure}

\begin{figure}[ht]
\begin{center}
\includegraphics[scale=0.60]{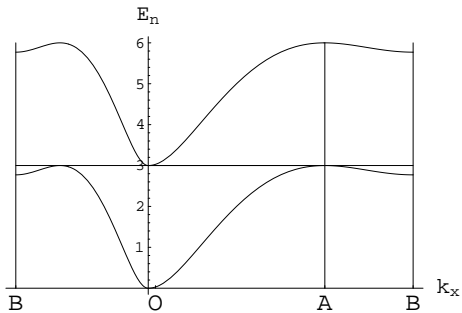}
\vspace{.2cm} \caption{The three dice lattice energy bands along
the main directions in ${\bf{k}}$-space. Special points in the
hexagonal Brillouin zone are O: origin, A: vertex, B: midpoint of
an edge.}
\label{dicebands.fig}
\end{center}
\end{figure}

Our simplest example is the dice (or T3) lattice, Fig.~\ref{dice.fig}, which
has a three-site unit cell for which Eq.~(1) is easily
diagonalized in linear spin wave theory. This is a system with a
net magnetic moment per unit cell and therefore it has a Goldstone
mode with a quadratic dispersion as in a ferromagnet. For
completeness, we present an outline of the spin wave calculation
for this simple case. The magnetic and the structural cell unit
cells are the same and have three sites, one on sublattice A with
$z=6$, and two on sublattice B, both with $z=3$. One introduces
Holstein-Primakoff operators of three types,
corresponding to each of these sites: $a_i, b_i$,$c_i, (i=1,...,N$
where $N$ is the number of unit cells) along with their adjoints,
obeying the appropriate bosonic commutation relations. Introducing
Fourier-transformed operators $a_{\vec{k}} =
\frac{1}{\sqrt{N}}\sum e^{i{\vec{k}} {\vec{r_i}}} a_i$,
$b(c)_{\vec{k}} = \frac{1}{\sqrt{N}}\sum e^{-i{\vec{k}}
{\vec{r_i}}} b(c)_i$, the linearized Hamiltonian is $H_{lin} = -
JS(S+1)N_b + JS H^{(2)}$, with $N_b = 2N$, where
\begin{eqnarray}
 H^{(2)} = \sum ( a^\dag_{\vec{k}}, b_{\vec{k}}, c_{\vec{k}}) \left(
\begin{array}{rrr}
6 & A & A^*\\
A^* & 3& 0\\
A & 0 & 3 \\
\end{array}
\right) \left(
\begin{array}{r}
a_{\vec{k}} \\
b^\dag_{\vec{k}} \\
c^\dag_{\vec{k}}\\
\end{array}\right). \label{diceham}
\end{eqnarray}
Here, $A(k_x,k_y) = e^{-i\sqrt{3}k_x/2}(
e^{i3k_y/2}+2\cos(\sqrt{3}k_x/2))$ ($k_{x},k_{y}$ 
are expressed in
dimensionless units). This Hamiltonian is diagonalized by a
generalization of the standard  case as described in \cite{white}
in order to find a set of bosonic operators $\alpha_{\vec{k}}$,
$\beta_{\vec{k}}$, $\gamma_{\vec{k}}$ such that
\begin{eqnarray}
 H^{(2)} = \sum ( \alpha^\dag_{\vec{k}}, \beta_{\vec{k}}, \gamma_{\vec{k}}) \left(
\begin{array}{rrr}
\omega_1(\vec{k}) & 0 & 0\\
0 &\omega_2(\vec{k}) & 0\\
0 & 0 & \omega_3(\vec{k}) \\
\end{array}
\right) \left(
\begin{array}{r}
\alpha_{\vec{k}} \\
\beta^\dag_{\vec{k}} \\
\gamma^\dag_{\vec{k}}\\
\end{array}\right). \label{diceham2}
\end{eqnarray}
An analytic solution is easily found and the results for the
variations of $\omega_i$ along different symmetry directions are
shown in Fig.~\ref{dicebands.fig}. The flat branch $\omega_3=3$ corresponds to an
excitation that lives on a subset of sites, situated on hexagons
of B-sites (as illustrated in Fig.~\ref{dice.fig}). In this linearized model,
each of the six spins around the central site precesses in
anti-phase with its neighbors. These states arise from the local
topology, which gives rise, similarly, to the ``Aharanov-Bohm
cages" discussed in \cite{vidal} for electrons subjected to a
magnetic flux on the dice lattice. The ground state energy per
site is $E_0/N_s \approx -0.633$ within linear spin wave theory,
compared to the quantum Monte Carlo result of
$-0.6384(1)$.

The local staggered magnetization of A-sites, $m_A = S -
 \langle a_i^\dag a_i\rangle$, and that of the B-sites, $m_B$, can be calculated
from the transformation matrices $O_{\vec{k}}$ that take the
set $\vec{a}_{\vec{k}}, b^\dag_{\vec{k}},c^\dag_{\vec{k}}$ into
the set
$\vec{\alpha}_{\vec{k}},\beta^\dag_{\vec{k}},\gamma^\dag_{\vec{k}}$.
One finds 

\begin{eqnarray}
m_A 
& = &
\frac{1}{2} - \frac{1}{N}\sum_{\vec{k}}
\frac{3 - \omega_2(\vec{k})}{3 + 2\omega_2(\vec{k})} \approx 0.356, \\
\nonumber 
m_B 
&=& \frac{1}{2} -\frac{1}{N}\sum_{\vec{k}}
\frac{3 - \omega_2(\vec{k})}{6 + 4\omega_2(\vec{k})} \approx 0.428.
\end{eqnarray}

The corresponding
quantum Monte Carlo results are
$0.3754(3)$, and $0.4381(2)$, respectively. Note that the quantum correction
to the classical B-sublattice magnetization, $\delta m_B=S-m_B$ is half that of
the A-sublattice correction $\delta m_A=S-m_A$. The local magnetization on the dice lattice thus
follows the trend seen on the octagonal tiling, being smaller on
sites with higher coordination number. As for the bond strengths,
there is only one type of nearest neighbor bond in this system, $b = \vert\langle
\vec{S_A} \vec{S_B}\rangle\vert = 0.316$ in linear spin wave
theory (quantum Monte Carlo result: $0.3189(1)$).

\subsection{\label{sec:level2 }III.B The CaVO lattice}

\begin{figure}[ht]
\begin{center}
\includegraphics[scale=0.5]{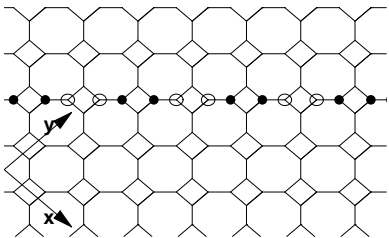}
\vspace{.2cm} \caption{The CaVO lattice, showing a state localized
on a strip (open circles correspond to bigger amplitudes than the
filled circles).}
\label{cavo.fig}
\end{center}
\end{figure}

\begin{figure}[ht]
\begin{center}
\includegraphics[scale=0.60]{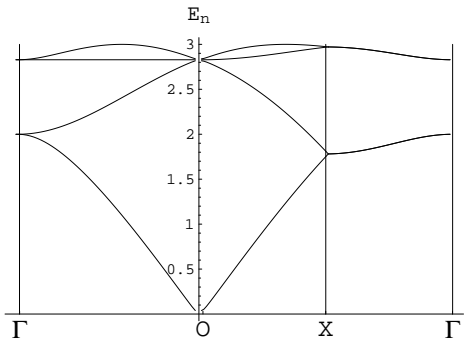}
\vspace{.2cm} \caption{The CaVO energy bands along the main
symmetry directions. Special points in the square
Brillouin zone are O: origin, X: midpoint of edge, $\Gamma$: vertex.}
\label{cavobands.fig}
\end{center}
\end{figure}

The second lattice considered here has four sites per unit cell,
and twice that in the magnetic unit cell. This is the T11
or CaVO lattice (after the calcium vanadium oxide compound whose spins lie
on the topological equivalent of this structure)
discussed previously (c.f. discussion in Ref.~\cite{richt}). We
introduce a numbering from 1 through 4 for the four sites of a
structural unit cell. The magnetic unit cell is doubled, with $x$ and $y$
axes rotated by $45^\circ$ with respect to the original axes as 
shown in Fig.~\ref{cavo.fig}.
We thus introduce the boson destruction operators
$a_{1-4}\ (b_{1-4})$ corresponding to sublattice A (B). 
The magnetic moment per magnetic unit cell is zero. The
transformation using bosonic operators, linearization, and Fourier
transformation for the sets of $a_i$ and $b_i$ are carried out
similarly as for the dice lattice. For a system of $N$ magnetic
unit cells, one finds $H_{lin} = -JS(S+1)N_b + JSH^{(2)}$, with
$N_b = 12N$ and
\begin{eqnarray}
 H^{(2)} = \left(
\begin{array}{r|r}
H_1 & H_2\\
\hline
H_2^\dag & H_1\\
\end{array}
\right), \label{ham}
\end{eqnarray}
where $H_1 = z_i \delta_{ij}=3 \delta_{ij}$, $H_2^\dag $
is the adjoint of $H_2$ with
\begin{eqnarray}
 H_2 = \left(
\begin{array}{rrrr}
1 & z^*_y & z^*_x& 0\\
1 & 1 & 0 & 1 \\
1 & 0 & 1 & 1 \\
0 & z_x & z_y & 1\\
\end{array}
\right),
\end{eqnarray}
where $z_{\mu} = \exp^{ik_\mu}$. The basis vectors are $\vec{a}^T
= (a_1,a_2,a_3,a_4,b_1^\dag,b_2^\dag,b_3^\dag,b_4^\dag)$.
Numerical diagonalization of this Hamiltonian leads to
$\vec{\alpha}_{\vec{k}} = O_{\vec{k}} \vec{a}_{\vec{k}}$, after
discretizing the Brillouin zone. We obtain four branches
$\omega_i$ ($i=1,..,4$) shown in Fig.~\ref{cavobands.fig}. There are dispersionless
directions, $k_x=\pm k_y$ for excitations living on strips, such as
the one shown in Fig.~\ref{cavo.fig}, of energy $E=2\sqrt{2}$. The participating sites occur with larger (smaller)
amplitudes as indicated by filled (open) circles.

The ground state energy per site for this
system within linear spin wave theory, obtained by a discrete
sum over the Brillouin zone, is $E_0/N_s \approx -0.5376$ , in
agreement with the value of Ueda et al.~\cite{ueda}. Ueda et al.
also show that in this system, the N\'eel state corrected by spin
waves has an energy very close to the energy of a plaquette
resonating valence bond (PRVB) state. The ground state energy
obtained in linear spin wave theory is seen to compare less well
with the quantum Monte Carlo results, $-0.55367(2)$ than in the
case of the dice lattice. Similar deviations between linear spin
wave theory and quantum Monte Carlo results are also obtained for
the local quantities: All sites in the CaVO lattice have the same
value of the staggered magnetization, which is obtained within 
linear spin wave theory as $m_i = S - \langle a_i^\dag a_i \rangle
= S - \langle b_i^\dag b_i \rangle \equiv m$. Our
k-space sum extrapolates to $m \approx 0.213$, close to the
value, 0.212, published in \cite{ueda}, but significantly larger
than the quantum Monte Carlo result, $0.1805(2)$. The increased
relevance of quantum fluctuations, which are not captured
appropriately within linear spin wave theory stems from the
closeness of the system to a quantum critical point~\cite{troyer},
beyond which the system enters a plaquette phase
with dominant singlet formations on the squares in
Fig.~\ref{cavo.fig}. 
This 
is reflected in the enhanced bond strength on the plaquettes, $b_{p}=
0.40731(6)$ compared to the remaining (dimer) bonds $b_{d}=0.2926(1)$.
Linear spin wave theory does not fully 
capture this effect, yielding
$b_{p} = |-S^2+ S(1 - 2 m + 2 Re \langle a_2^\dag b_4^\dag
\rangle | = 0.365$, and $b_{d} = |-S^2+ S(1
- 2 m + 2 Re \langle a_1^\dag b_1^\dag \rangle | = 0.344$, respectively.

\subsection{\label{sec:level3c}  III.C The crown lattice}

\begin{figure}[ht]
\begin{center}
\includegraphics[scale=0.4]{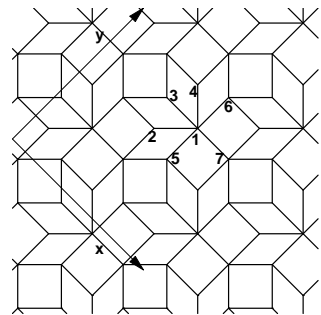}
\vspace{.2cm} \caption{ A unit cell of the crown lattice, with sites numbered from 1 to 7.}
\label{crownunit.fig}
\end{center}
\end{figure}

\begin{figure}[ht]
\begin{center}
\includegraphics[scale=0.50]{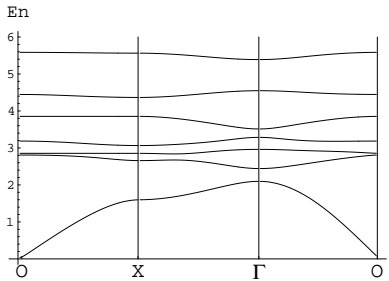}
\vspace{.2cm} \caption{ The seven crown lattice energy bands along some of the main
symmetry directions. Special points in the square
Brillouin zone are O: origin, X: zone boundary along 
$k_x$ axis, $\Gamma_1$: the $\pi,\pi$ vertex.}
\label{crownbands.fig}
\end{center}
\end{figure}

This is a square lattice system with seven sites per unit cell, labeled as
shown in Fig.~\ref{crownunit.fig}, and
twice this number in the antiferromagnetic
unit cell. Each unit has a mirror symmetry that reduces the number
of distinct environments to four. This periodic structure has
most, though not all, of the local environments present in the
quasiperiodic octagonal tiling, and can in fact be obtained via
projection from a four dimensional cubic lattice using the method
for obtaining approximants of the tiling outlined in \cite{oguey}.
Each of the
sites occurs once on each sublattice, as for the CaVO lattice, and
we introduce accordingly the Bogoliubov operators $a_i, a_i^\dag$
for sites of sublattice A and $b_i, b_i^\dag$ for sites of
sublattice B ($i=1,...,7$). The linearized Hamiltonian is 
$H_{lin} = -JS(S+1)N_b + JSH^{(2)}$, where $N_b = 14N$. $H^{(2)}$
takes the form given in Eq.~({\ref{ham}}) in a basis $\vec{a}^T =
(a_1,a_2,..,a_7,b_1^\dag,b_2^\dag,..,b_7^\dag)$ where
\begin{eqnarray}
(H_1)_{ij} = z_i\delta_{ij}; \vec{z} = (6,3,3,3,4,4,5),
\end{eqnarray}
and
\begin{eqnarray}
 H_2 = \left(
\begin{array}{rrrrrrr}
0 & \overline{z}_y\overline{z}_x&\overline{z}_y\overline{z}_x& \overline{z}_x &\overline{z}_y\overline{z}_x&1&\overline{z}_y\\
1&0&0&0&0 & 1 & \overline{z}_y \\
1 & 0 &0&0&\overline{z}_x& 1 & 0 \\
1 & 0 &0&0&\overline{z}_x& 0 & 1 \\
1 & 0 &\overline{z}_y &1&0&0&\overline{z}_y \\
1 & 1 &1&0&0 & 0 &1 \\
1&1&0&1&1 & z_x & 0\\
\end{array}
\right).
\end{eqnarray}
The $14 \times 14$ system is diagonalized to find the new basis
set in terms of operators ${\alpha_i, \beta_i}$ as in the
preceding cases. The solution was obtained by discretizing
$\vec{k}$ in the Brillouin zone of the crystal, whose unit cell is defined
in a coordinate system oriented at
$45^\circ$ with respect to the original axes as shown in Fig.~\ref{crownunit.fig}.
The seven distinct energy levels
thus obtained are plotted in Fig.~\ref{crownbands.fig} for some of the main directions in
$\vec{k}$-space.

\begin{figure}[ht]
\begin{center}
\includegraphics[scale=0.40]{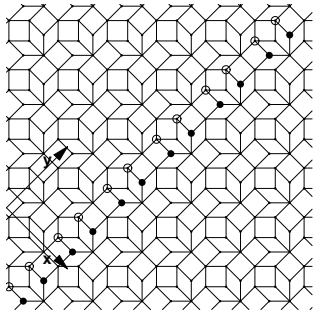}
\vspace{.2cm} \caption{ States on strips (see text). Amplitudes are different on
the sites with open (filled) circles.}
\label{crownstrip.fig}
\end{center}
\end{figure}

There are no localized states, however there $is$ a dispersionless
direction, $k_y=0$, for excitations located on ribbons (or
strips) which are localized in the $x$ but extended in the $y$
direction. As shown in Fig.~\ref{crownstrip.fig}, the sites with non-zero amplitudes
lie on the corners of rows of rectangles (formed by the double
squares). Each rectangle has a pair of sites of $z=3$, and a pair
of sites of $z=4$, whose amplitudes are different (corresponding
to open (closed) circles). This state corresponds to the energy
band $E\approx 3.85$ which is, as can be seen in Fig.~\ref{crownbands.fig},
flat in the $OX$ direction.

\begin{figure}[ht]
\begin{center}
\includegraphics[width=7cm]{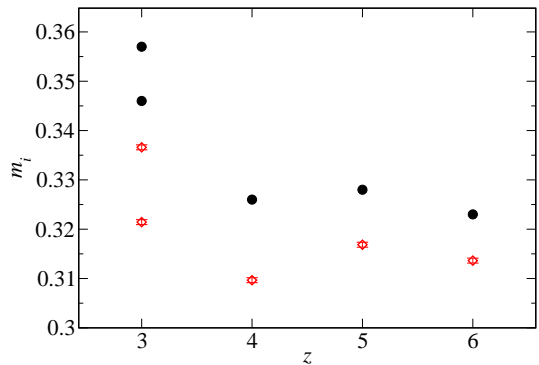}
\vspace{.2cm} \caption{
Dependence of the local staggered magnetization $m_i$ on the
coordination number z for the inequivalent sites of the crown
lattice. Quantum Monte Carlo data (open) are compared to linear spin
wave theory results (full). }
\label{compare_m.fig}
\end{center}
\end{figure}

\begin{figure}[ht]
\begin{center}
\includegraphics[width=7cm]{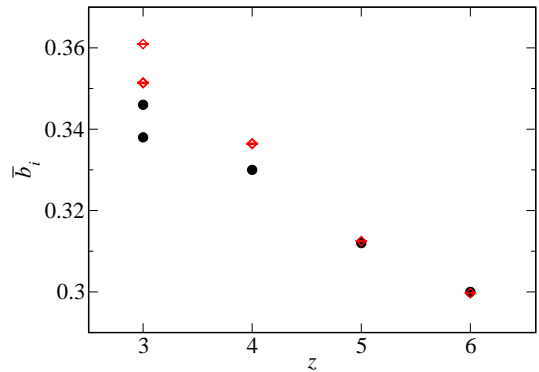}
\vspace{.2cm} \caption{
Dependence of the local averaged bond strength $\overline{b}_i$ on the coordination number z for
the inequivalent sites of the crown lattice. Quantum Monte Carlo data (open circles) are compared to
linear spin wave theory results (full circles).
}
\label{compare_b.fig}
\end{center}
\end{figure}

The ground state energy per site for this system is found within spin wave theory to be 
$E_0/N_s = -0.6475$, close to the value obtained in spin wave theory for the quasiperiodic octagonal tiling, of 
$E_0/N_s = -0.646$ \cite{wess2}. 
The quantum Monte Carlo results are $-0.6602(1)$ for the crown lattice and $=-0.6581(1)$ for the
octagonal tiling~\cite{wess3}, respectively.

The linear spin wave theory results for the local staggered
magnetization are shown in Fig.~\ref{compare_m.fig} for different
values of the coordination number $z$ of the sites, and compared to
the quantum Monte Carlo results.  We find that linear spin wave theory
does rather well in describing this inhomogeneous system.  Apart from
an overall overestimation of the antiferromagnetic order, both the
spin wave theory and quantum Monte Carlo results exhibit similar
qualitative features: for $z=3$ there are two different values of the
staggered magnetization---the sites labeled 2 and 4 have a larger
staggered magnetization than the site labeled 3 (more is said about
this below).  The case of $z=4$ is particularly interesting, since
these sites are found by quantum Monte Carlo to have the $smallest$
value of $m_i$.  As a result one finds a non-monotonic
dependence of the local staggered magnetizations on $z$
(comparing $z=4$ to $z=5$).
Such a
non-monotonic dependence of the local staggered magnetization on $z$
was noticed in the quantum Monte Carlo~\cite{wess1} and numerical spin
wave calculations for the octagonal tiling~\cite{wess2}. We will
return to this phenomenon in the following section.

For the crown lattice, the number of bonds per structural unit cell is 14,
but only 8 are inequivalent due to the
mirror symmetry.
From their respective strengths, we obtain the values of the
averaged local bond strength shown as a function of $z$ in
Fig.~\ref{compare_b.fig}, where the results from linear spin wave
theory are compared to the quantum Monte Carlo data. We again find
a difference between the two types of $z=3$ sites. Furthermore, we
observe a monotonic decrease of $b_i$ with the coordination number
$z$. This indicates that correlations beyond the nearest-neighbor distance 
are important for the non-monotonic behavior of the local
staggered magnetization $m_i$ seen in Fig.~\ref{compare_m.fig}.


\section{\label{sec:level4} IV. Heisenberg stars}

In this section, we present a semiclassical theory for Heisenberg
stars, which reconciles our finding of a monotonically decreasing
$m_i(z)$ with the intuition that increasingly highly coordinated
lattices (in $d\geq2$) should approach a saturated order
parameter.

\begin{figure}[ht]
\begin{center}
\includegraphics[width=3cm]{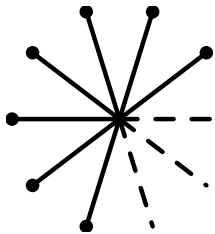}
\caption{A Heisenberg star: a central spin is coupled to $\zeta$
mutually un-coupled neighbors.} \label{hstar.fig}
\end{center}
\end{figure}

A Heisenberg star, studied in detail from a different perspective
in Ref.~\onlinecite{richterstar}, consists of a central site
coupled to $\zeta$ neighbors, which are mutually disconnected 
(Fig.~\ref{hstar.fig}).  We denote the destruction
operators of the Holstein-Primakoff bosons for the central spin,
${\vec s}$, with $a$, and those for the outer spins, ${\vec S}_i$,
with $B_i$, $i=1\ldots\zeta$. We then have \be H=-\zeta S^2+S \siz
a^\dag a+B_i^\dag B_i+aB_i+a^\dag B_i^\dag\ . \label{eq:hp} \ee
Next, we carry out a unitary transformation, $U$, between the
$b_i$: $b_i=U_{ij}B_j$, with $U_{1i}=1/\sqrt{\zeta}$, so that \bea
H&=&-\zeta S^2+\\ \nn &S&\left[\zeta a^\dag a+b_1^\dag b_1+
\sqrt{\zeta}(ab_1+a^\dag b_1^\dag)
+\sum_{i=2}^\zeta b_i^\dag b_i\right] \ . \eea All modes $b_i$ with
$i\geq2$ correspond to precessions of outer spins around the
central spin so that their net transverse component vanishes --
the analogue of the local modes for the dice lattice mentioned
above.

The lowering of the magnetization is due to the anomalous terms
linking $b_1$ with $a$. This part of the problem can be diagonalised
with a standard Bogoliubov transformation matrix
\be\nn
O=\left(
{\cosh\theta \ \sinh\theta}\atop{\sinh\theta\ \cosh\theta}
\right)\ ,
\ee
with
\be\nn
\tanh(2\theta)=2\sqrt{\zeta}/(\zeta+1)\ .
\ee
 This yields
for the magnetization of the central and outer spins, respectively,
\bea\label{eq:mstar}
m&=&S-1/(\zeta-1)
\\ \nn
M&=&S-1/[\zeta(\zeta-1)] .
\eea
Note the following two features. Firstly, as $\zeta$ grows, the
quantum reduction to the sublattice magnetization decreases,
keeping with the idea that a high exchange field implies a reduction in
the importance of quantum fluctuations. Secondly, however, it is
systematically the highly coordinated central site which has the lower
magnetization: at large $\zeta$, the central $m$ is reduced by
$1/\zeta$, whereas the outer $M$ are each only reduced by $1/\zeta^2$.

The origin of this phenomenon can be traced to the structure of the
Hamiltonian (Eq.~\ref{eq:hp}): Holstein-Primakoff bosons in a
bipartite antiferromagnet are created only in pairs, one each on
opposite ends of each bond. The higher a site's coordination, the
more anomalous terms in the Hamiltonian act to
create such boson pairs. As the number of bosons quantifies the reduction
in $m_i$, high coordination thus
implies low magnetization.  This is
reminiscent of the order-by-disorder effect encountered in a simple
hopping problem, where one finds that in the ground state,
particles are more likely to be found near highly coordinated sites.
At the same time, a high exchange field (corresponding to large
diagonal terms in the Hamiltonian) makes such bosons costly and
suppresses their overall number -- which is why high-$z$ lattices
have a higher $m$ than low-$z$ ones.

The above order-by-disorder mechanism is operative only 
when differently coordinated
sites are located at opposite ends of a bond. If, for the sake of 
argument,
one were to consider an inhomogeneous system formed by patching
together large pieces of uniform lattices (say,  square and
honeycomb), the local $m$ in the interior of each piece
would of course reflect that of the relevant parent lattice -- and
in this case, $m(z)$ would {\em increase} with $z$.
In other words, as placing a boson on a site involves placing a boson
on one of a neighboring sites as well, $m(z)$ is of course not a
purely local quantity. The next-best {\em guess} would be that, for
sites with the same $z$, the magnetization will be lowest for those
sites whose neighbors have the lowest average coordination. This is in
keeping with the result that $0.3366(5) =m_2 > m_3= 0.3214(5)$
for the crown lattice and also with the bimodal splitting of the local staggered magnetization
observed for the five-fold coordinated sites on to quasiperiodic octagonal tiling~\cite{wess1,wess2}.
In other words, this analysis shows that the monotonicity in
$m_i(z)$ is a trend, not a law, in that details of the mode spectrum
can in fact violate it.

We close this section with two side remarks. Firstly, note that
the Heisenberg star for $\zeta>1$ has a non-vanishing total
magnetization, ${\cal M}$, which has enabled us to obtain a finite
reduction of $m$ due to quantum fluctuations, even though it is a
0-dimensional system. For $\zeta=1$, when the the total
magnetization vanishes, the corrections (Eq.~\ref{eq:mstar})
themselves diverge. Secondly, since ${\cal M}$ is invariant
under the action of the Hamiltonian, it follows that
$S-m=\zeta(S-M)$. By the same token, simply dressing the N\'eel
state on the dice lattice leads to $S-m_{A}=2(S-m_{B})$, a
result which holds not only for linear spin waves but also 
in our quantum Monte Carlo simulations.

\section{\label{sec:level5} V. Discussion and Conclusions}
We have presented results for local properties of bipartite
inhomogeneous quantum magnets. They lead us to conclude that in
periodic systems containing inequivalent sites, there is in
general a trend towards decreased values of $m_i$ and
$\overline{b}_i$ with increasing local coordination number.  This
trend was also previously observed in a quasiperiodic system.  The
results for the crown lattice, in particular, show a qualitative
similarity with the results that were earlier obtained for the
self-similar quasiperiodic octagonal tiling.  We have provided a
simple analytical model system, the Heisenberg star, the
semiclassical solution of which explains this qualitative
behavior.  
Whereas it is well known that quantum fluctuations can lead to a range
of ordering phenomena in otherwise disordered magnets~\cite{hassan},
it is interesting to ask whether a novel {\em dis}ordering effect can
be obtained by increasing $z$ for a sublattice, as suggested by a
decreasing $m_i(z)$. However, as we have shown above, an increase of $z$
on one sublattice will generically make the ordering {\em more} robust
everywhere, although not uniformly so for all
sublattices. This reconciles in a simple way the observed decrease of
$m_i(z)$ with increasing $z$ with the naive expectation that the
ordered moment $m_i$ should grow with the coordination of the lattice.

SW acknowledges LPS, Orsay and CEA Saclay for hospitality during this collaboration.
The numerical calculations were performed at HRLS Stuttgart and NIC J\"ulich.

\end{document}